\documentclass[twocolumn,prl,aps,superscriptaddress]{revtex4-1}

\usepackage{array} 
\usepackage{paralist} 
\usepackage{verbatim} 
\usepackage{helvet}
\usepackage{amssymb}
\usepackage{mathrsfs}
\usepackage{graphicx}
\usepackage{dsfont}
\usepackage{subfigure}
\usepackage[sumlimits]{amsmath}
\usepackage{braket} 
\usepackage{nicefrac} 
\usepackage{accents} 
\usepackage{bbm} 
\usepackage{subfigure}
\usepackage{textcomp} 
\usepackage{booktabs} 

\newcommand{\e}{\ensuremath{\mathrm{e}}}
\newcommand{\n}{\ensuremath{\mathrm{n}}}

\newcommand{\Tr}{\operatorname{Tr}}
\newcommand{\ident}{\mathds{1}}

\newcommand{\Hhf}{H_{\textrm{HF}}}
\newcommand{\HD}{H_{\textrm{D}}}
\newcommand{\Hmw}{\ensuremath{H_{\text{\textmu w}}}}
\newcommand{\omegaD}{\omega_{\textrm{D}}}

\newcommand{\HDp}{\ensuremath{\HD^{\shortparallel}}}
\newcommand{\HDc}{\ensuremath{\HD^{\times}}}

\newcommand{\mdsh}{\mbox{-}}

\newlength{\dhatheight}
\newcommand{\hhatDS}[1]{%
    \settoheight{\dhatheight}{\ensuremath{\hat{#1}}}%
    \addtolength{\dhatheight}{-0.35ex}%
    \hat{\vphantom{\rule{1pt}{\dhatheight}}%
    \smash{\hat{#1}}}}
\newcommand{\hhatTS}[1]{%
    \settoheight{\dhatheight}{\ensuremath{\hat{#1}}}%
    \addtolength{\dhatheight}{-0.35ex}%
    \hat{\vphantom{\rule{1pt}{\dhatheight}}%
    \smash{\hat{#1}}}}    
\newcommand{\hhatS}[1]{%
    \settoheight{\dhatheight}{\ensuremath{\scriptstyle{\hat{#1}}}}%
    \addtolength{\dhatheight}{-0.175ex}%
    \hat{\vphantom{\rule{1pt}{\dhatheight}}%
    \smash{\hat{#1}}}}
\newcommand{\hhatSS}[1]{%
    \settoheight{\dhatheight}{\ensuremath{\scriptscriptstyle{\hat{#1}}}}%
    \addtolength{\dhatheight}{-0.07ex}%
    \hat{\vphantom{\rule{1pt}{\dhatheight}}%
    \smash{\hat{#1}}}}
\newcommand{\hhat}[1]{\mathchoice{\hhatDS{#1}}{\hhatTS{#1}}{\hhatS{#1}}{\hhatSS{#1}}}

\begin{document}

\title{Parallel Information Transfer in a Multi-Node Quantum Information Processor}
\author{T. W. Borneman}
\affiliation{Department of Nuclear Science and Engineering, Massachusetts Institute of Technology, Cambridge, MA, USA}
\affiliation{Institute for Quantum Computing, Waterloo, ON, Canada}
\author{C. E. Granade}
\affiliation{Institute for Quantum Computing, Waterloo, ON, Canada}
\affiliation{Department of Physics, University of Waterloo, Waterloo, ON, Canada}
\author{D. G. Cory}
\affiliation{Department of Nuclear Science and Engineering, Massachusetts Institute of Technology, Cambridge, MA, USA}
\affiliation{Institute for Quantum Computing, Waterloo, ON, Canada}
\affiliation{Perimeter Institute for Theoretical Physics, Waterloo, ON, Canada}
\affiliation{Department of Chemistry, University of Waterloo, Waterloo, ON, Canada}

\begin{abstract}
We describe a method for coupling disjoint quantum bits (qubits) in different local processing nodes of a distributed node quantum information processor. An effective channel for information transfer between nodes is obtained by moving the system into an interaction frame where all pairs of cross-node qubits are effectively coupled via an exchange interaction between actuator elements of each node. All control is achieved via actuator-only modulation, leading to fast implementations of a universal set of internode quantum gates. The method is expected to be nearly independent of actuator decoherence and may be made insensitive to experimental variations of system parameters by appropriate design of control sequences. We show, in particular, how the induced cross-node coupling channel may be used to swap the complete quantum states of the local processors in parallel.


\end{abstract}

\maketitle

A distributed, multi-node structure has been suggested as a convenient means of arranging qubits in an experimentally realizable quantum computer architecture \cite{mehring_spinbus_2006, suter_endohedral_2002, twamley_quantum_2003, lloyd_qc_1993, kane_silicon-based_1998, kielpinski_architecture_2002, jiang_distributed_2007,cappellaro_coherence_2009}. Such a structure requires the ability to define an array of disjoint quantum processors that may be controlled locally, with communication between local processors provided by a coupling between nodes that may be turned on or off in turn. One method of satisfying these requirements is to append a small number of qubits to each node of an array of nearest-neighbor coupled actuator elements. The actuator elements provide local control of the surrounding processor qubits and a means of transferring information between nodes \cite{bermudez_electron_2011}. In this work, we present a method for generating a universal set of gates between any pair of disjoint cross-node qubits via isotropic actuator couplings. 

An effective cross-node processor coupling network is created by taking advantage of four-body coupling terms between actuator and processor elements that appear in a manifold of excited states unused for quantum information storage. By moving into an appropriate interaction frame, the four-body coupling terms appear as two-body couplings between every pair of cross-node processor qubits in a properly defined computational manifold. While this complete cross-node coupling network allows for a computationally universal set of operations between nodes, we present an explicit implementation of a parallel swap of the complete quantum mechanical states of two local quantum processors. Consideration of this representative entangling operation serves to motivate the broader applicability of the induced information transfer channel. Additionally, since information is never explicitly stored for an appreciable amount of time on the actuators -- which are exposed to higher levels of noise than the processor elements -- we expect the channel to be nearly independent of actuator decoherence.
\begin{figure}
 \centering
 \includegraphics[width=8cm]{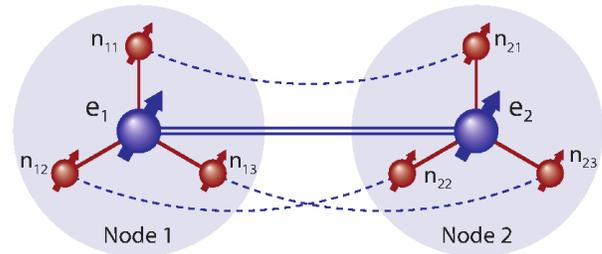}
 \caption{\textbf{2$\times$(1e-3n) Node Schematic.} The nodes are taken to be identical, with resolved anisotropic hyperfine interactions (solid red lines) between electron actuator spins and nuclear processor spins. The local processors are initially disjoint, but may be effectively coupled (dotted lines) by modulating an isotropic actuator exchange interaction (solid blue double line) and moving into an appropriate microwave Hamiltonian interaction frame. The spin labeling is e$_i$ for electron actuator spins and n$_{ij}$ for nuclear processor spins, where $i$ labels the nodes and $j$ labels the qubits.}
 \label{fig:name-conventions}
\end{figure}

The details of the method will be discussed from the standpoint of a spin-based, distributed node quantum information processor. This system contains all the necessary physics and is representative of many other modalities being considered for experimental realizations of quantum information processing. For example, Rydberg atom excitations of neutral atoms \cite{jaksch_fast_2000,saffman_quantum_2010}, inductive coupling of superconducting qubits \cite{majer_coupling_2007,majer_spectroscopy_2005}, and Bloch wave dispersion in cavity devices \cite{duan_colloquium:_2010,soderberg_phonon-mediated_2010} all take the form of an isotropic dipolar coupling. Direct dipolar interactions also naturally occur in spin-based devices such as semiconductor quantum dots \cite{loss_quantum_1998,petta_coherent_2005}, silicon-based devices \cite{kane_silicon-based_1998,morton_solid_2008}, nitrogen-vacancy defect centers in diamond \cite{wrachtrup_processing_2006}, and other solid-state spin systems \cite{moussa_solidnmr_2006,suter_scalable_2002,harneit_fullerene_2002}. The methods developed in this work may be readily extended to these systems. 

In our spin-based model, each node consists of a single actuator electron spin coupled via resolved anisotropic hyperfine interactions to each of $k$ qubits of a local nuclear spin processor (Fig. \ref{fig:name-conventions}). Control over the local processors is achieved via electron-only modulation \cite{hoffman_primary_1995}, taking advantage of the relative strength of the hyperfine interaction to generate a universal set of fast quantum gates on the nuclear spins \cite{hodges_universal_2008}, which serve as excellent storage elements for quantum information due to their relatively long coherence times \cite{suter_spinqubits_2008}. The internode coupling of actuators is given by an isotropic dipolar or exchange interaction between electrons. The spatial separation of the nodes is taken to be sufficient for any cross-node dipolar interactions of nuclear spins to be negligible. 

\begin{figure}
 \includegraphics[width=8cm]{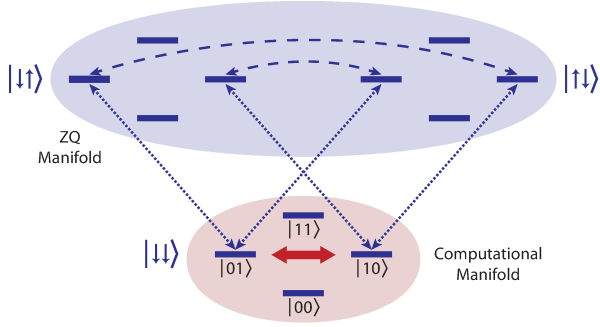}
 \caption{\textbf{Energy Structure of 2$\times$(1e-1n) System.} Quantum information is encoded in the states $\ket{\downarrow\downarrow00}$, $\ket{\downarrow\downarrow01}$, $\ket{\downarrow\downarrow10}$, and $\ket{\downarrow\downarrow11}$ of the actuator ground-state (Computational) manifold, where arrows indicate electron actuator spin states and binary digits indicate nuclear processor spin states. The desired transition (bold red arrow) for a {\scshape swap} operation is implemented by applying a selective microwave field that induces transitions between the two manifolds (dotted lines), effectively moving the induced cross-node processor transitions (dashed lines) in the actuator zero-quantum (ZQ) manifold to the Computational manifold. Note that the actuator excited state manifold, $\ket{\uparrow\uparrow}$, is not included as it is not involved in the internode transfer process.}    
 \label{fig:EnergyLevelDiagram}
\end{figure}

The state structure of a two node system with one electron actuator spin and one nuclear processor spin each -- a 2$\times$(1e-1n) system -- is shown in Fig. \ref{fig:EnergyLevelDiagram}. The computational basis states of the local quantum processors are defined in the ground-state manifold of the actuators. This choice of encoding allows us to implement gates between the disjoint processors by taking advantage of an induced cross-node coupling in the zero-quantum (ZQ) manifold of actuator excited states not used for information storage. 
\newcommand{\HZ}{\ensuremath{H_{\text{Z}}}}

We derive the form of the cross-node coupling for a general 2$\times$(1e-$k$n) system. The nodes are taken to be identical with energy structure given by a dominant, quantizing electron Zeeman interaction, $\HZ^\e$, with a strong static magnetic field oriented along the laboratory $\hat{z}$ direction; a corresponding nuclear Zeeman interaction, $\HZ^{\n}$; an anisotropic hyperfine interaction between electron and nuclear spins, $\Hhf^{\e\mdsh \n}$; and a dipolar interaction between electron spins, $\HD^{\e\mdsh \e}$:
\begin{equation}
H^{2\e \mdsh k\n} = \HZ^\e + \HZ^\n + \Hhf^{\e\mdsh\n} + \HD^{\e\mdsh\e}.
\label{eq:2e2nHamiltonian}
\end{equation}
In a frame rotating at the electron Zeeman frequency, the resulting secular Hamiltonians are given in terms of the usual spin-$\frac{1}{2}$ Pauli operators as \cite{abragam_principles_1983}
\begin{equation}
\begin{split}
& \HZ^\e = \sum_k{\omega_z^{k}\left(\sigma_z^{\n_{1k}}+\sigma_z^{\n_{2k}}\right)} \\
& \HD^{\e\mdsh\e} = \omega_d \left( 2\sigma_z^{\e_1}\sigma_z^{\e_2} - \sigma_x^{\e_1}\sigma_x^{\e_2} - \sigma_y^{\e_1}\sigma_y^{\e_2}  \right) \\
& \Hhf^{\e\mdsh\n} = \sum_k{\vec{A}^k \cdot \left( \sigma_z^{\e_1} \vec{\sigma}^{\n_{1k}} + \sigma_z^{\e_2} \vec{\sigma}^{\n_{2k}} \right)},
\end{split}
\label{eq:Hamiltonians}
\end{equation}
where the vectors $\vec{A}^{k} = A_x^k \hat{x} + A_y^k \hat{y} + A_z^k \hat{z}$ represent the strengths and directions of the hyperfine coupling between the $k$th nuclear spin in each node and the corresponding actuator, $\omega_z^k$ is the strength of the nuclear Zeeman interaction for the $k$th nuclear spin, $\omega_d$ is the strength of the dipolar interaction, and $\vec{\sigma} = \sigma_x \hat{x} + \sigma_y \hat{y} + \sigma_z \hat{z}$.

The nuclear spins are quantized in an effective field given by the vector sum of the hyperfine and nuclear Zeeman interactions. The resulting eigenstates are non-commuting, allowing for universal control of the nuclear spins via electron-only control \cite{hodges_universal_2008}. Note that, since the nodes are identical, the above Hamiltonians do not provide the ability to selectively address nodes. Universal control over the entire 2$\times$(1e-$k$n) system is obtained by adding a term to the Hamiltonian that spatially labels the nodes to allow for local operations. These terms are not included in the present discussion as they are not necessary for the implementation of gates between nodes, and may be effectively turned off. We only require that the differences in the hyperfine coupling strengths within each node are large enough for each pair of identical spins to be spectroscopically resolved. This requirement limits the number of qubits per node \cite{cappellaro_coherence_2009} but, due to the inherent inefficiency of designing control sequences for a large number of particles, it is advantageous to keep the size of nodes small and rely on the ability to swap qubit states between nodes to implement large-scale quantum algorithms. 

\begin{figure}
 \centering
 \includegraphics[width=6.75cm]{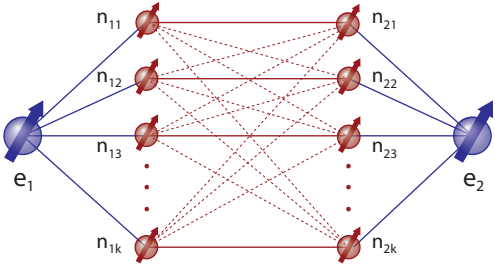}
 \caption{\textbf{Internode Coupling Network.} The induced coupling network of local processor elements consists of interactions between every pair of cross-node spins. For implementation of a parallel swap operation, we wish to keep interactions between identical spins (solid red lines) while refocusing all other interactions (dotted lines).}
 \label{fig:MultipleSpinsSchematic}
\end{figure}

The full set of interactions accessible by evolution under the Hamiltonians in (\ref{eq:Hamiltonians}) is given by the Lie algebra generated by taking Lie brackets to all orders \cite{ramakrishna_controllability_1995,schirmer_complete_2001}. In particular, the second-order bracket, $\left[[\HD^{\e\mdsh\e},\Hhf^{\e\mdsh\n}],\Hhf^{\e\mdsh\n}\right]$, takes the form of a four-body inter-node interaction, given by an effective cross-node nuclear spin dipolar coupling, $\HD^{\n\mdsh\n}$, along with flip-flop transitions of the electron spins:
\begin{equation}
[[\HD^{\e\mdsh\e}, \Hhf^{\e\mdsh\n}], \Hhf^{\e\mdsh\n}] \propto \omega_d \left(\sigma_+^{\e_1} \sigma_-^{\e_2} + \sigma_-^{\e_1} \sigma_+^{\e_2}\right) \otimes \HD^{\n \mdsh\n}.
\label{eq:Commutator}
\end{equation}
The resulting nuclear spin dynamics in the ZQ manifold may be decomposed into a sum of coupling terms, $\HD^{\ell m}$, that act on every pair of cross-node spins, $\n_{1\ell}$ and $\n_{2m}$ (Fig. \ref{fig:MultipleSpinsSchematic}). Each $\HD^{\ell m}$ may be written in terms of the well-known dipolar alphabet, with $\sigma_{\pm} = \sigma_x \pm i \sigma_y$ \cite{abragam_principles_1983}:
\begin{subequations}
\begin{align}
 \tilde{A}_{\ell m} & = A_z^{\ell} A_z^m \left(\sigma_z^{\n_{1\ell}} \sigma_z^{\n_{2m}} \right) \\
 \tilde{B}_{\ell m} & = \left(A_x^\ell A_x^m + A_y^\ell A_y^m \right) \left(\sigma_+^{\n_{1\ell}} \sigma_-^{\n_{2m}} + \sigma_-^{n_{1\ell}} \sigma_+^{n_{2m}} \right) \\
 \nonumber & +\left(A_x^\ell A_y^m - A_y^\ell A_x^m \right) \left(\sigma_+^{\n_{1\ell}} \sigma_-^{\n_{2m}} - \sigma_-^{\n_{1\ell}} \sigma_+^{\n_{2m}} \right) \\
 \tilde{C}_{\ell m} & =\left(A_x^\ell A_z^m - i A_y^\ell A_z^m \right)\left(\sigma_+^{\n_{1\ell}} \sigma_z^{\n_{2m}} + \sigma_z^{\n_{1\ell}} \sigma_+^{\n_{2m}} \right) \\
 \nonumber & + \left(A_z^\ell A_x^m - A_z^\ell A_z^m - i A_z^\ell A_y^m + i A_y^\ell A_z^m \right) \sigma_z^{\n_{1\ell}} \sigma_+^{\n_{2m}} \\
 \tilde{E}_{\ell m} &= \left( A_x^\ell A_x^m - A_y^\ell A_y^m - i A_x^\ell A_y^m - i A_y^\ell A_x^m \right) \sigma_+^{\n_{1\ell}} \sigma_+^{\n_{2m}} \\
 \tilde{D}_{\ell m} &= \tilde{C}_{\ell m}^{\dag} , \tilde{F}_{\ell m} = \tilde{E}_{\ell m}^{\dag}.
\end{align}
\end{subequations}

After application of an appropriate microwave control field, these four-body interactions in the ZQ manifold appear as effective cross-node two-body couplings in the computational manifold. To demonstrate this, we consider the implementation of a particularly powerful operation: a parallel swap of entire local processor states between nodes at once. For the case of a single nuclear spin per node, the relevant ZQ transitions are $\ket{\uparrow\downarrow01}\bra{\downarrow\uparrow10}$ and $\ket{\uparrow\downarrow10}\bra{\downarrow\uparrow01}$. Application of a microwave field with matrix elements $\ket{\uparrow\downarrow}\bra{\downarrow\downarrow}$ and $\ket{\downarrow\uparrow}\bra{\downarrow\downarrow}$ of strength commensurate with the ZQ transitions transforms both of the transitions to a swap operation in the computational manifold: $\ket{\downarrow\downarrow01}\bra{\downarrow\downarrow10}$ (See Fig. \ref{fig:EnergyLevelDiagram}).

When multiple nuclear spins are present in each node, a parallel swap operation requires suppressing couplings between non-identical spins ($\ell \neq m$) while retaining couplings between identical spins ($\ell=m$). This may be accomplished by exploiting the difference in symmetry between the prefactors of the coupling operators for identical versus non-identical spins.

Consider, as an example, a 2$\times$(1e-2n) system. The effective Hamiltonians of the induced interactions may be written as $\HDp + \HDc$, where $\HDp = \HD^{11} + \HD^{22}$ and $\HDc = \HD^{12} + \HD^{21}$. The effective dipolar coupling strength for $\HDc$ appears as odd order in $\vec{A}^1$ and $\vec{A}^2$, while each term in $\HDp$ appears as even order. Thus, by inverting the state of only the second (or first) spin in each node halfway through free evolution under the induced Hamiltonians, we can generate a zeroth-order average Hamiltonian of only the desired $\HDp$ interactions \cite{haeberlen_coherent_1968}. Higher order terms in the average Hamiltonian may be suppressed through the use of more sophisticated pulses or by applying the evolution-pulse-evolution cycle at a rate fast compared to $\omegaD$. This symmetry argument may be easily generalized to a larger number of nuclear spins per node by applying a binomially expanding set of inversion pulses to properly select the desired couplings \cite{jones_efficient_1999}.

We now consider how to isolate the desired interaction (\ref{eq:Commutator}) from other elements of the algebra. One method of suppressing the extraneous terms is to use a composite pulse sequence to generate an effective Hamiltonian for which the desired second order commutator is the dominant term. Concretely, recall that by the BCH expansion, $e^X e^Y = \exp(X + Y + \frac12[X, Y] + \frac1{12} [X,[X,Y]] + \frac1{12} [Y, [Y,X]] + \cdots)$. By recursively applying this expansion, we can derive an identity that suppresses all terms below second order:
\begin{equation}
 \label{eq:bch-pulse}
 e^X e^Y e^{-X} e^{-Y} e^{-X} e^Y e^{X} e^{-Y} = e^{[X, [X, Y]] + \cdots}.
\end{equation}
By making the correspondence $X=\Hhf^{\e\mdsh\n}$ and $Y=\HD^{\e\mdsh\e}$, we obtain a pulse composed of sequential periods of only electron dipolar or hyperfine evolution, leading to the effective propagator,
\begin{equation}
  U(8\tau) \approx e^{i\tau^3 [[\HD^{\e\mdsh\e},\Hhf^{\e\mdsh\n}],\Hhf^{\e\mdsh\n}]},
\end{equation}
where higher-order terms have been neglected. We may also suppress the undesired terms by numerically optimizing experimentally robust microwave pulses which achieve the desired interaction while suppressing all other interactions \cite{fortunato_design_2002,cory_nmr_2000,khaneja_optimal_2005,borneman_octcpmg_2010}.

A final consideration is the sensitivity of the induced channel to actuator noise processes. We claim that by never transferring complete qubit state information to the actuators, we may operate in a regime where any portion of the information present in the ZQ manifold arrives back to the computational manifold before it is corrupted. We may quantitatively determine the robustness of the channel to actuator noise by comparing, as a function of noise strength, the Hilbert-Schmidt inner product fidelity, $F(\hhat{S}_{\text{ideal}}, \hhat{S}_{\text{noisy}}) = \Tr(\hhat{S}_{\text{ideal}}^\dagger \hhat{S}_{\text{noisy}})/d^2$, between $d^2$ dimensional superoperators representing the channel in the presence of noise, $\hhat{S}_{\text{noisy}}$, and in the ideal noiseless case, $\hhat{S}_{\text{ideal}}$. 

\begin{figure}
      \includegraphics[width=0.8\linewidth]{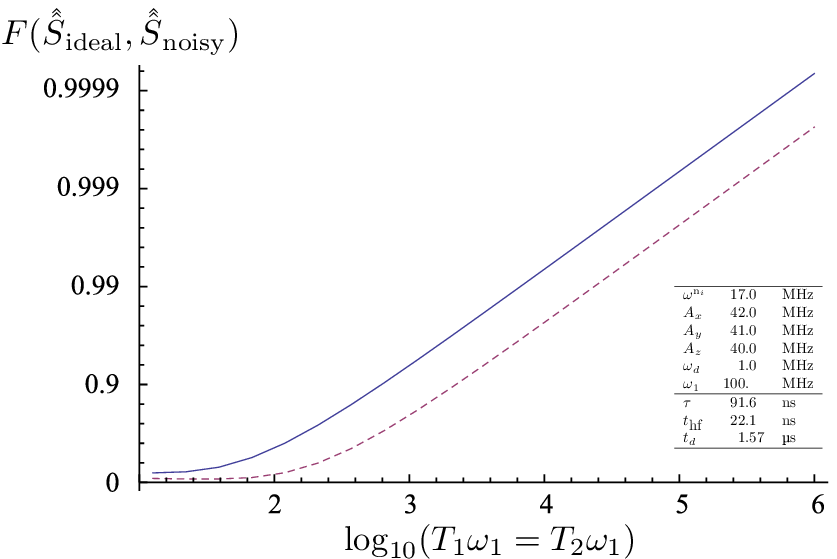}
    \caption{\textbf{Channel Noise Robustness.} A plot of the channel fidelity, $F$, as a function of noise strength, $T_1 = T_2$, scaled by computing $-\log[1-F]$. The induced channel (solid line) performs significantly better than a serial swap operation (dotted line) \cite{supp_mat}. Given a modest Rabi frequency, $\omega_1 = 100$ MHz, the induced channel is nearly independent of actuator decoherence for electron relaxation times above 100 $\mu$s.}
    \label{fig:robustness}
\end{figure}

The ideal channel is generated by a Liouvillian operator, $\hhat{L}$, corresponding to unitary evolution only. The noisy channel includes two dissipation operators, $\hhat{D}_1$ and $\hhat{D}_2$, describing the relaxation of $e_1$ and $e_2$, respectively: 
\begin{equation}
 \hhat{S}_{\text{noisy}}(t) = e^{-i t\hhat{L} + t \hhat{D}_1 + t \hhat{D}_2}.
\end{equation}
A physically motivated model of noise is a contribution of phase and amplitude damping applied seperately to each electron, which leads to a dissipator,
\begin{eqnarray}
  \hhat{D} = && -\frac12 \left(\Gamma_1 + \Gamma_2\right) \left(E_- \otimes \ident + \ident \otimes E_- \right) \\*
  \nonumber  && \hphantom{\Gamma_1} + \Gamma_1 \sigma_+ \otimes \sigma_+ + \Gamma_2 E_- \otimes E_-,
\end{eqnarray}
where $E_{+} = \ket{0}\bra{0}$ and $E_{-} = \ket{1}\bra{1}$ are projection operators. The noise strength is parameterized by $\Gamma_{1,2}$, which are related to the commonly used energy, $T_1$, and coherence, $T_2$, relaxation times by $\Gamma_1 = 1 / {T_1}$ and $\Gamma_2 = \left(2 T_1 - T_2\right) / \left(T_1 T_2\right)$. A plot of the superoperator fidelity versus noise strength is shown in Fig. \ref{fig:robustness}. The noise has a minimal effect on the operation of the channel for values of $T_1 \omega_1 \gtrsim 10^4$. Assuming a modest Rabi frequency of $\omega_1$ = 100 MHz, actuator relaxation times of 100 $\mu$s are required to avoid significant corruption of the information during transfer. Currently achievable relaxation times for electron spins are well within this range \cite{morton_solid_2008,heidebrecht_quantum_2006}.

By taking advantage of the additional degrees of freedom present in an actuator based system, and manipulating the naturally occuring actuator interactions between nodes, we were able to create an effective channel between initially disjoint local processors that allows the parallel transfer of $k$-qubit states between nodes, effectively independent of actuator decoherence. The channel takes the form of four-body cross-node interactions in the zero-quantum manifold of actuator states which, after moving into an appropriate microwave interaction frame, appear as effective two-body couplings of cross-node qubits in the computational manifold. The resulting complete cross-node coupling network may be used to generate a universal set of gate operations between nodes. We expect the techniques described in this work to be applicable to a wide-variety of quantum devices, with minimal need for modification.

\textit{Acknowledgements} - This work was supported by the Canadian Excellence Research Chairs (CERC) Program and the Canadian Institute for Advanced Research (CIFAR).

\newcommand{\swapgt}{\ensuremath{\textsc{swap}}}
\newcommand{\T}{\ensuremath{\mathrm{T}}}

\onecolumngrid
\newpage

In this supplemental material, we elaborate on the calculations used in the main text to justify the robustness of our proposed wide quantum channel against relaxation processes acting on the electron actuators. In particular, we compare the pulse sequence  for isolating the second-order commutator of two operators given in Equation (5) of the main text to a sequence of pulses implementing serial $\swapgt$ gates. For simplicity, we shall assume that there is one nuclear spin in each node, and shall hence drop the node index from our nuclear spin labels.

Concretely, by alternately suppressing and inverting the hyperfine and dipolar couplings, we can implement a unitary $U$ generated by the effective cross-node nuclear dipolar coupling for a cycle of length $t_c = 8\tau$,
\begin{align*}
 U(8\tau) & = e^{-i\tau \Hhf^{\e\mdsh \e}} e^{-i\tau \HD^{\e\mdsh \n}} e^{+i\tau \Hhf^{\e\mdsh \e}} e^{+i\tau \HD^{\e\mdsh \n}} e^{+i\tau \Hhf^{\e\mdsh \e}} e^{-i\tau \HD^{\e\mdsh \n}} e^{-i\tau \Hhf^{\e\mdsh \e}} e^{+i\tau \HD^{\e\mdsh \n}} \\
 & \approx e^{+i\tau^3 \left[\Hhf^{\e\mdsh\n}, \left[\Hhf^{\e\mdsh\n}, \HD^{\e\mdsh\e}\right]\right]} \\
 & = \exp\left(+i\tau^3 \omega_d (\sigma_+^{\e_1} \sigma_-^{\e_2} + \sigma_-^{\e_1}\sigma_+^{\e_2} ) \otimes \HD^{\n\mdsh\n}\right).
\end{align*}

A reasonable choice for $t_c$, and hence for $\tau$, is such that the nuclear spins evolve for a phase $\pi / 2$ under the action of the effective nuclear dipolar interaction. We can estimate this by choosing $\tau$ such that
\[
 \tau^3 \left\Vert \HD^{\e\mdsh\e} \right\Vert \left\Vert \Hhf^{\e\mdsh\n} \right\Vert^2 = \pi/2.
\]

\newcommand{\Lhf}{\ensuremath{\hhat{L}_{\textrm{HF}}}}
\newcommand{\LD}{\ensuremath{\hhat{L}_{\textrm{D}}}}
\newcommand{\Lmw}{\ensuremath{\hhat{L}_{\text{\textmu w}}}}

\newcommand{\Sprep}{\ensuremath{\hhat{S}_{\text{prep}}}}
\newcommand{\Strace}{\ensuremath{\hhat{S}_{\text{trace}}}}
\newcommand{\Swqci}{\ensuremath{\hhat{S}_{\text{wqc, ideal}}}}
\newcommand{\Swqcn}{\ensuremath{\hhat{S}_{\text{wqc, noisy}}}}
\newcommand{\Sswapi}{\ensuremath{\hhat{S}_{\text{\swapgt, ideal}}}}
\newcommand{\Sswapn}{\ensuremath{\hhat{S}_{\text{\swapgt, noisy}}}}
Since we are only interested in how this pulse sequence acts on the states of the nuclear spins, we compose the unitary evolution with superoperators representing state preparation  $\Sprep[\rho^{\n_1\n_2}] = \Ket{\downarrow_{e_1} \downarrow_{e_2}} \Bra{\downarrow_{e_1} \downarrow_{e_2}} \otimes \rho^{\n_1\n_2}$ and partial tracing $\Strace[\rho^{\e_1\e_2\n_1\n_2}] = \Tr_{\e_1\e_2} \left(\rho^{\e_1\e_2\n_1\n_2}\right)$ of the electron spins. We also include a selective microwave Hamiltonian to cause mixing between the computational and zero quantum manifolds, as illustrated in Figure 2 of the main text:
\[
 \Hmw = \omega_1\left(\sigma_x^{\e_1}+\sigma_x^{\e_2}\right)\otimes\left(E_+^{\n_1}E_-^{\n_2}+E_-^{\n_1}E_+^{\n_2}\right),
\]
where $E_+^{\n_1}E_-^{\n_2} = \ket{01}\bra{01}$ and $E_-^{\n_1}E_+^{\n_2} = \ket{10}\bra{10}$ are projection operators on the nuclear spins. The action of the ideal wide quantum channel is given by
\begin{align*}
 \Swqci & = 
  \Strace \cdot \left(
    e^{-i\tau \Lhf^{\e\mdsh \e} + i\tau \Lmw}\ 
    e^{-i\tau \LD^{\e\mdsh \n} + i\tau \Lmw}\ 
    e^{+i\tau \Lhf^{\e\mdsh \e} + i\tau \Lmw}\ 
    e^{+i\tau \LD^{\e\mdsh \n} + i\tau \Lmw}\  \right. \\ & \qquad \left.
    e^{+i\tau \Lhf^{\e\mdsh \e} + i\tau \Lmw}\ 
    e^{-i\tau \LD^{\e\mdsh \n} + i\tau \Lmw}\ 
    e^{-i\tau \Lhf^{\e\mdsh \e} + i\tau \Lmw}\ 
    e^{+i\tau \LD^{\e\mdsh \n} + i\tau \Lmw}
  \right) \cdot \Sprep,
\end{align*}
where $\LD$, $\Lhf$ and $\Lmw$ are the Liouvillian representations of the respective Hamiltonians $\HD$, $\Hhf$ and $\Hmw$. In the column-stacking basis for Liouville space, we can write that
\[
 \hhat{L} = \ident\otimes H - H^\T \otimes \ident = \ident\otimes H - H^* \otimes \ident,
\]
where $H^*$ is the complex conjugate of $H$, so that $\hhat{L}[\rho] = H \rho - \rho H = \left[H, \rho\right]$ represents the Liouville-von Neumann equation $\partial_t \rho = -i\hhat{L}[\rho]$. 

To model how decoherence affects this sequence, we act on the electrons with the dissipator given in Equation (8) of the main body. This dissipator is then added to the generator for each interval in the above pulse sequence, so that
\begin{align*}
 \Swqcn & = 
  \Strace \cdot \left(
    e^{-i\tau \Lhf^{\e\mdsh \e} + i\tau \Lmw + \tau \hhat{D}}\ 
    e^{-i\tau \LD^{\e\mdsh \n} + i\tau \Lmw + \tau \hhat{D}}\ 
    e^{+i\tau \Lhf^{\e\mdsh \e} + i\tau \Lmw + \tau \hhat{D}}\ 
    e^{+i\tau \LD^{\e\mdsh \n} + i\tau \Lmw + \tau \hhat{D}}\  \right. \\ & \qquad \left.
    e^{+i\tau \Lhf^{\e\mdsh \e} + i\tau \Lmw + \tau \hhat{D}}\ 
    e^{-i\tau \LD^{\e\mdsh \n} + i\tau \Lmw + \tau \hhat{D}}\ 
    e^{-i\tau \Lhf^{\e\mdsh \e} + i\tau \Lmw + \tau \hhat{D}}\ 
    e^{+i\tau \LD^{\e\mdsh \n} + i\tau \Lmw + \tau \hhat{D}}
  \right) \cdot \Sprep.
\end{align*}

We are then concerned with how robust the action of our wide quantum channel is against the relaxation; that is, we are concerned with how accurately we may model our channel on the nuclear spins as decoherence-free. To answer this, we consider the process fidelity given by the Hilbert-Schmidt inner product between the ideal and noisy superoperators, $F(\Swqci, \Swqcn) = \Tr(\Swqci^\dagger \cdot \Swqcn) / d^2$, where $d$ is the dimension of the Hilbert space of nuclear spin states.

For comparison to a serial $\swapgt$ protocol, we consider a composition of serial $\swapgt$ gates
\[
 \swapgt^{\n_1\n_2} = \left(\swapgt^{\n_1\e_1} \swapgt^{\n_2\e_2} \right) \cdot \swapgt^{\e_1\e_2} \cdot \left(\swapgt^{\n_1\e_1} \swapgt^{\n_2\e_2} \right),
\]
where we have grouped in parentheses those $\swapgt$ gates which can be performed in parallel. By noting that for two spins, $\swapgt = \exp\left[\frac{\pi}{2} (\frac12 \vec\sigma^1\cdot\vec\sigma^2 - \ident^1\ident^2)\right]$, where $\vec\sigma^i$ is the vector of Pauli matrices on spin $i$, we can represent $\swapgt^{\n_1\n_2}$ by the pulse sequence
\begin{align*}
 \swapgt^{\n_1\n_2} = &
    \exp\left(t_{\text{hf}} \omega_{\text{hf}} \left[
      \frac12 \vec\sigma^{\e_1} \cdot \vec\sigma^{\n_1} - \ident^{\e_1} \ident^{\n_1} +
      \frac12 \vec\sigma^{\e_2} \cdot \vec\sigma^{\n_2} - \ident^{\e_2} \ident^{\n_2}
    \right] \right) \cdot \\ &
    \exp\left(t_d \omega_d \left[
      \frac12 \vec\sigma^{\e_1} \cdot \vec\sigma^{\e_2} - \ident^{\e_1} \ident^{\e_2}
    \right] \right) \cdot \\&
    \exp\left(t_{\text{hf}} \omega_{\text{hf}} \left[
      \frac12 \vec\sigma^{\e_1} \cdot \vec\sigma^{\n_1} - \ident^{\e_1} \ident^{\n_1} +
      \frac12 \vec\sigma^{\e_2} \cdot \vec\sigma^{\n_2} - \ident^{\e_2} \ident^{\n_2}
    \right] \right),
\end{align*}
where $\omega_{\text{hf}} = \left\Vert\Hhf^{\e\mdsh\n}\right\Vert$, and where $t_{\text{hf}}$ and $t_d$ are chosen to ensure that $t_{\text{hf}} \omega_{\text{hf}} = t_d \omega_d = \pi / 2$. As before, we can represent this sequence as an action on the nuclear spins alone by representing each Hamiltonian as a Liouvillian, calculating the superoperators and composing with the preparation and trace steps $\Sprep$ and $\Strace$. We model decoherence in the same way as before, adding the dissipator $\hhat{D}$ to each step of this serial protocol. Doing so, we obtain superoperators $\Sswapi$ and $\Sswapn$ for the ideal and noisy serial $\swapgt$ protocols, respectively.

\begin{figure}
    \includegraphics[width=0.75\linewidth]{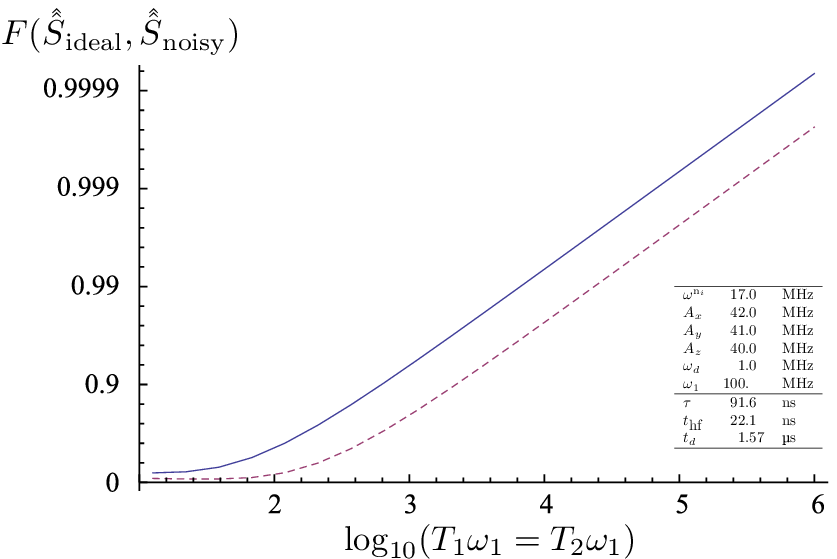}
    \caption{\textbf{Channel Noise Robustness.} A plot of the channel fidelity $F$ as a function of noise strength, $T_1 = T_2$, scaled by computing $-\log[1-F]$. The solid curve shows the fidelity of the wide quantum channel, while the dotted curve shows the fidelity of a serial $\swapgt$ protocol. Values of Hamiltonian parameters used for simulation are included in the table inset. For electron relaxation times above 100 $\mu$s, the channel is nearly independent of actuator decoherence.
}
    \label{fig:robustness2}
\end{figure}  

In Figure 4 of the main text, reproduced here as Figure \ref{fig:robustness2}, we compare the fidelities $F(\Swqci, \Swqcn)$ and $F(\Sswapi, \Sswapn)$, observing that the wide quantum channel performs significantly better against electron decoherence than the serial $\swapgt$ protocol.

\end{document}